\documentclass[10pt]{article}
\usepackage{color}
\usepackage{epsfig}
\usepackage{epstopdf}
\usepackage{amsmath,amssymb}
\usepackage{txfonts}

\begin{document}

\title{Spectrally engineered broadband photon source for two-photon quantum interferometry }

\author{Abu Thomas$^{1,*}$,  Mackenzie Van Camp$^{1}$, Olga Minaeva$^{2}$, David Simon$^{3}$\\ and Alexander V. Sergienko$^{1,4}$}
\maketitle
\begin{center}

$^1$Dept. of Electrical and Computer Engineering \& Photonics Center, Boston University, 8 Saint Mary's St., Boston, MA 02215, USA
\\ 
$^2$Dept. of Biomedical Engineering, Boston University, 44 Cummington St., Boston, MA 02215, USA
\\
$^3$Dept. of Physics and Astronomy, Stonehill College, 320 Washington Street, Easton, MA 02357, USA
\\
$^4$Dept. of Physics, Boston University, 590 Commonwealth Ave., Boston, MA 02215, USA \\
$^*$Corresponding author: abthomas@bu.edu 
\end{center}


\begin{abstract}

We present a new approach to engineering broadband sources of entangled photon pairs for quantum interferometry. The source is based on quasi-phase-matched spontaneous parametric down conversion in a titanium diffused periodically poled lithium niobate waveguide with a strongly-chirped poling period. The proposed non-standard asymmetric poling  mitigates phase distortions associated with the process of chirping. Asymmetric poling significantly broadens the entangled source bandwidth while preserving high visibility quantum interferometric sensing.
\end{abstract}


\section{Introduction}

Quantum interference with entangled-photon states is a major tool in quantum information processing, communication, and quantum metrology, dating to the early days of quantum optics \cite{HOM1987}.  Polarization entanglement is a common method for encoding qubits and for quantum information processing \cite{Mattle96,Bouwmeester97,Zeilinger98,Tittel98}, while spectral entanglement is at the heart of the dispersion cancellation effect used in quantum measurement schemes  \cite{Franson92,Steinberg92_PRA,Kwiat93,Abouraddy02_PRA}. The simultaneous entanglement of multiple variables is often called hyper-entanglement \cite{Atature01_PRA,Kwiat-hyperentangled2005}. Frequency-polarization hyper-entanglement gave rise to a special type of quantum interferometer \cite{SergienkoJOSAB1995} used in quantum metrology for super-resolution sensors to measure polarization-dependent optical delays\cite{Dauler99,fraine2012evaluation}.

The central element of any high-resolution quantum interference measurement device is an ultra-broadband source of polarization and frequency entangled photon pairs. The process of type-II spontaneous parametric down conversion (SPDC) in nonlinear crystals is a preferred source of hyper-entangled states~\cite{Kwiat95PRL,martin2010polarization}, but unfortunately, the natural spectral bandwidth of emitted pairs in type-II interaction is rather narrow. The bandwidth is always smaller than that of type-I SPDC because the different dispersive properties of  orthogonally-polarized down-converted photons tighten the requirements for phase matching.  Early demonstrations of broadband type-II sources simply used very thin (50-100 microns) non-linear crystals, taking advantage of the fact that the phase-matching bandwidth   in the nonlinear medium is inversely proportional to the interaction length \cite{Dauler99}. However, the resulting small interaction volume inevitably lead to a very low quantum state production rate, limiting its practicality.

The majority of recent entangled-photon source implementations use quasi-phase-matched (QPM) nonlinear crystals in order to avoid cumbersome arrangements in bulk crystal. The ideology of quasi-phase-matching proposed in the early days of nonlinear optics \cite{Armstrong62,FejerIEEE1992} blossomed into a mainstream technology  \cite{Myers95, SilberhornPPLNheralded}. It enables researchers to achieve high-quality phase matching  without need for rotating the bulk nonlinear crystal searching for the correct angle. Quasi-phase-matching enables effective nonlinear conversion over much longer interaction regions when compared with bulk crystal phase matching, helping achieve significantly higher production rates of entangled photon pairs. However, the increase in the nonlinear interaction length also narrows the down conversion spectral bandwidth. The technique of chirping (changing the periodicity of the modulation pattern) is a tool that can broaden the nonlinear conversion spectrum. Chirping has been successfully implemented in conventional broadband parametric amplifiers and converters. One known problem is that linear chirping of periodic structure in nonlinear crystals inevitably deforms the initial $\left(sin(x)/x\right)$ spectral phase matching profile  \cite{FejerChirpJOSAB2008,FraineSource2012} and reduces parametric  amplification of Gaussian signals. A special process of chirping pattern apodization was introduced with the goal of smoothing the spectral intensity profile in parametric amplifiers \cite{Huang2006,HeeseOptExpress2012} leading to a quasi-Gaussian spectral intensity envelope.

The production of significantly broadband ($>$100 nm wide) and simultaneously high quality hyper-entangled photon states for quantum interferometry is a more demanding task. In addition to the broad and smooth spectral intensity profile of the source, quantum interferometry relies on the purity of the generated spectral entanglement. It is known from the early days of quantum optics that the quality of quantum interference depends on complete indistinguishability of contributions from photons produced in different parts of the nonlinear interaction region, and on the stable phase relationship between all conjugated signal and idler spectral components \cite{Rubin94,SergienkoJOSAB1995}. The effect of indistinguishability can be described using a spectral amplitude and phase distribution in the output signal. A convenient way to model the SPDC in a QPM structure is to consider a sequence of coherently pumped conversion processes in each nonlinear domain. The photons generated in different domains interfere at the output of the crystal and produce the final broadband spectrum. The timing of each photon's arrival at the surface of detectors in quantum interferometry is defined by the  location of its origin inside the nonlinear source. In the standard case of a coherent pump and {\em uniform} poling period the difference in arrival of two photons is a linear function of this location and is directly connected with the differential group velocity delay accumulated inside the source.  One polarization emerges ahead of the other because of birefringence. This constant phase dependance, together with a symmetric spectral intensity distribution, opens up a means to compensate for the increased distinguishability due to difference in the arrival time. Compensation is achieved by using a linear external polarization delay \cite{Kwiat95PRL,Rubin94} in order to reinstate indistinguishability of contributions from orthogonally-polarized photon pairs generated in the front and in the back portion of the crystal.

The situation changes in the case of a strongly chirped QPM crystal. The chirp-induced  distortion of timing  between photons generated in different regions of the nonlinear material becomes so significant that it strongly deforms the interference pattern by creating beating-type modulations. The reduced visibility of the quantum interference pattern and its deformed shape are caused by the spectral intensity modulation and by the asymmetry of higher-order phase derivatives in the spectral profile of the down converted radiation. This detrimental phase-distortion effect must be mitigated in order to restore high-quality quantum interference and to build quantum interferometric sensors capable of resolving sub-femtosecond time delays \cite{fraine2012evaluation,fraine2011precise}.

This paper demonstrates that a special poling profile that is asymmetric in regards to the center of the physical sample in a strongly-chirped quasi-phase-matched nonlinear crystal can remedy this complex problem, enlarging the entangled photon source bandwidth while simultaneously achieving high visibility in a quantum interferometer.  The main novel feature is that the symmetrization and establishment of indistinguishability between signal and idler spectral contributions occurs in the {\em phase-amplitude parameter space} instead of traditional spatial symmetrization procedure. The positive  effect of the proposed symmetrization in phase-amplitude space is illustrated by numerical simulation of coincidence interferometry outcomes using quantum states produced with nonlinearly-chirped poling patterns imposed on titanium diffused lithium niobate waveguides (Ti:PPLN). The diffusion of titanium creates waveguides in lithium niobate that are capable of supporting type-II nonlinear interactions, in contrast to  proton-exchange embedded PPLN waveguides that can support only one polarization mode.

\section{Quantum interferometry with two-photon hyper-entangled quantum states}

\begin{figure}[h]
\centering\includegraphics[width=10	cm]{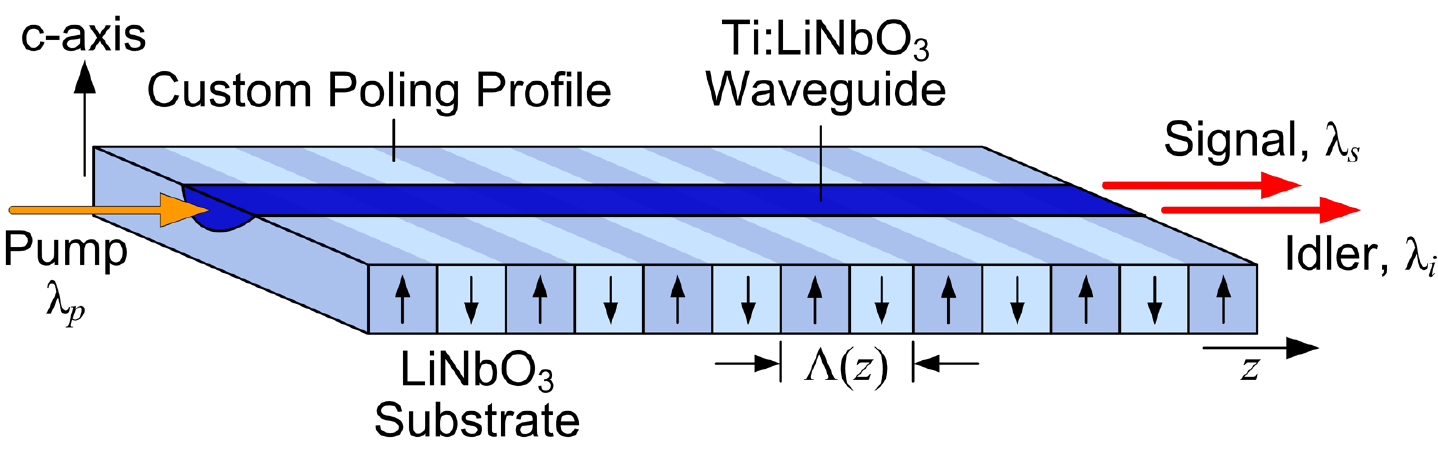}
\caption{Schematic of the type-II SPDC source based on periodically poled Ti:PPLN waveguide.}
\label{PPLN}
\end{figure}

We consider a nonlinear optical process of type-II SPDC  pumped with a monochromatic, plane-wave laser of angular frequency $\omega_p$ in a Ti-diffused waveguide in lithium niobate with a custom periodic poling pattern (see Fig.~\ref{PPLN}).

Using the well-established general theoretical description  \cite{Rubin94, fraine2011precise}, the two-photon portion of the outgoing state can be expressed (up to overall constants) as

\begin{align}
\left|\psi\right.\rangle&\sim\int\int~d\omega_s~d\omega_i~A(\omega_s,\omega_i)~\left|\omega_s\right.\rangle_H |\omega_i\rangle_V\\
A(\omega_s,\omega_i)&=\sum_{m=1}^M~(-1)^m~L_m\mbox{sinc}(\Delta \beta L_m/2)~e^{-i\Delta \beta (L_m/2+\sum_{n=m+1}^M L_n)},
\end{align}

where $A(\omega_s,\omega_i)$ is the two-photon amplitude of type-II SPDC as a function of signal frequency,~$\omega_{s}$, and idler frequency,~($\omega_{i}$), and the sinc function is defined by $\mbox{sinc}(x)=\sin(x)/x$. Here, $\Delta\beta=\beta_H(\omega_{s})+\beta_V(\omega_{i})+\beta_d-\beta_H(\omega_p)$ is the difference of propagation constants of interacting fields in the waveguide, and $\beta_d=\frac{2\pi}{\Lambda}$ where $\Lambda$ is a period of inverse domain poling (see Fig.\ref{PPLN}). $L_m$ is the length of the $m$-th domain in the structure. The subscripts $H$ and $V$ indicate horizontal and vertical polarizations respectively, and we have arbitrarily identified the horizontal polarization with the signal. The summation in equation (2) represents  a coherent superposition of contributions occurring in individual domains. The phase accumulated by passive linear propagation of photons created in each domain is shown as the summation in the exponent. Note that this model is capable of describing arbitrary complex poling profiles since there is no restriction on the length of each domain. The total length of the waveguide (or periodically poled interaction region) is $L_{WG}=\sum_{m=1}^{m=M}L_m$.

\begin{figure}[h]
\centering\includegraphics[width=10	cm]{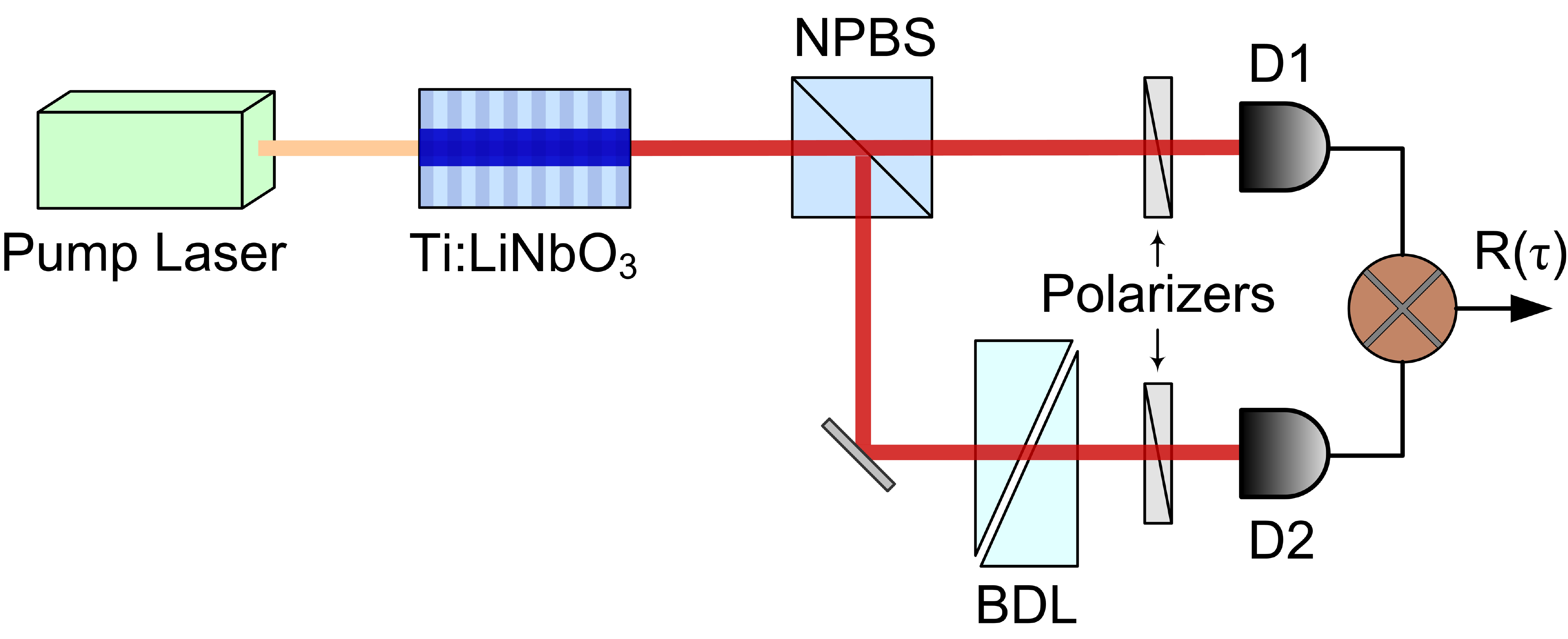}
\caption{Schematic of the quantum interferometer based on hyper-entangled states generated in type-II down conversion.}
\label{Setup}
\end{figure}

The state $|\psi\rangle$ produced by collinear type-II interaction in the waveguide is introduced into a  two-photon correlation interferometer (Fig.~\ref{Setup}). The polarization entangled state, $(|HV\rangle-|VH\rangle)/\sqrt{2}$, is post-selected by recording only coincidence counts between two single-photon detectors $D_1$ and $D_2$. The polarization interferometer consists of  a non-polarizing beam splitter~(NPBS) and a diagonally-oriented polarizer in each output arm of the beam splitter. Additionally a birefringent delay line~(BDL) made of two crystal quartz prisms to provide variable thickness is placed in the output arm of the NPBS leading to detector $D_2$. After projecting incoming photons onto a diagonal polarization basis using polarizers in both arms, we record  coincidence counts using $D_1$, $D_2$ and a time interval analyzer.

In the case of a monochromatic pump with angular frequency $\omega_p$, the signal~($\omega_s$) and idler ($\omega_i$) frequencies are anti-correlated. Combined with energy conservation, ($\omega_s+\omega_i=\omega_p$), this allows us to write the signal and idler frequencies in terms of a detuning frequency $\Omega$, \begin{equation} \omega_s=\omega_0+\Omega, \qquad \omega_i=\omega_0-\Omega ,\end{equation} where $\omega_0=\omega_p/2$. Note that $\Omega $ may be either positive or negative, so each photon can be either blue- or red- detuned from the degeneracy point $\omega_0$ .

The coincidence counting rate $R(y)$ as a function of the birefringent delay line thickness $y$ is

\begin{align}
R(y)=\int_{-\infty}^{\infty} d\Omega~\left| A(\omega_0+\Omega,\omega_0-\Omega )~\left(e^{ik_V(\omega_0-\Omega)y}~+~~e^{ik_H(\omega_0+\Omega)y}\right) \right|^2,
\label{Rz}
\end{align}

\noindent where $k_{H}$~($k_{V}$) is the horizontal~(vertical) propagation constant of the BDL. The two terms in the integrand represent the two amplitudes that contribute to coincidence counts~---~(i)~a horizontally-polarized signal photon at frequency $\omega_0+\Omega$ reaching D1 along with a vertical idler photon at $\omega_0-\Omega$ reaching D2, or (ii) a vertical idler at $\omega_0-\Omega$ reaching D1 along with a horizontal signal at $\omega_0+\Omega$ reaching D2. In each case, only the photon in the lower branch receives the extra birefringent phase shift. The two possibilities are made indistinguishable by diagonal projection in each arm, so that they may interfere.

By reversing the sign of the integration variable in the second term of the sum and then factoring out $e^{ik_H(\omega_0-\Omega)y}$ from both terms,  $R(y)$ in Eq.~(\ref{Rz})  may be put in the form:

\begin{align}
R(y) &= \int_{-\infty}^{\infty} d\Omega~\left| A(\omega_0+\Omega,\omega_0-\Omega )~e^{ik_V(\omega_0-\Omega)y}~+~~A(\omega_0-\Omega,\omega_0+\Omega )e^{ik_H(\omega_0-\Omega)y} \right|^2\\
&=\int_{-\infty}^{\infty}~d\Omega~\left|f(\Omega)~e^{i(k_V(\omega_0-\Omega)-k_H(\omega_0-\Omega))y}~+~f(-\Omega)\right|^2\\
&=\int_{-\infty}^{\infty}~d\Omega~\left|f(\Omega)~e^{i\theta}~+~f(-\Omega)\right|^2,
\label{Rz2}
\end{align}

where $f(\Omega)=A(\omega_0+\Omega,\omega_0-\Omega)$.
The phase difference in the crystal quartz birefringent delay line, $\theta(\Omega,y)~=~[k_V(\omega_0-\Omega)-k_H(\omega_0-\Omega)]y$, can be simplified by considering only the first two orders in the dispersive series $\theta(\Omega,y)=\omega_0\tau_{ph}-\Omega\tau_{gr}$. Higher dispersive orders in crystal quartz can be neglected. The minus sign is selected by the appropriate orientation of the BDL optical axis. Here $\tau_{ph}$ and $\tau_{gr}$ are the polarization phase and group delays in the BDL, respectively. For clarity, the phase delay after traveling a distance $y$ in the delay line is $\tau_{ph}=(y/\omega_0)[k_V(\omega_0)-k_H(\omega_0)]$ and the group delay is $\tau_{gr}=y~(dk_V/d\Omega-dk_H/d\Omega)$.

The amplitude of the down conversion signal $f(\Omega)=|f(\Omega)|e^{i\phi(\Omega)}$ is generally a complex function. $\left|f(\Omega)\right|^2$ is proportional to joint spectral intensity of the signal and idler waves, while the phase $\phi(\Omega)$  contains important information about the  time differences between photons in the pair at the exit face of the source. The final structure of the quantum interference pattern $R(y)$ is defined by the interplay of both the magnitude $\left|f(\Omega)\right|$ and phase $\phi(\Omega)$ contributions. In order to investigate how the phase of down conversion $\phi(\Omega)$ affects the coincidence counts $R(y)$ in the intensity interferometer, we employ a Taylor expansion around the zero detuning point $\Omega=0$, corresponding to the degenerate frequency of the phase-matching condition $\omega_s = \omega_i = { \omega_p}/2$. We obtain

\begin{align}
\phi(\Omega)&=\phi(0)+\frac{d\phi}{d\Omega}\Omega+\frac{1}{2!}\frac{d^2\phi}{d\Omega^2}\Omega^2+\frac{1}{3!}\frac{d^3\phi}{d\Omega^3}\Omega^3+...\\
&=\underbrace{\phi(0)+\frac{1}{2!}\frac{d^2\phi}{d\Omega^2}\Omega^2+...}_{even \ terms}+\underbrace{\frac{d\phi}{d\Omega}\Omega+\frac{1}{3!}\frac{d^3\phi}{d\Omega^3}\Omega^3+...}_{odd \ terms},
\end{align}

The reason for grouping even and odd order terms becomes clear when we introduce the phase  $\phi(\Omega)$ explicitly into the expression for coincidence counts Eq.~(\ref{Rz2}):

\begin{equation}
R(y)=\int_{-\infty}^{\infty}~d\Omega~\left|~|f(\Omega)|~e^{i\theta}~e^{i[\phi(\Omega)-\phi(-\Omega)]}+~|f(-\Omega)|~\right|^2
\end{equation}

The perfect anti-correlation between signal and idler frequencies forces the $\Omega$-dependent part of the phase term to be of the form $\Delta(\Omega)=\phi(\Omega)-\phi(-\Omega)$, so that \textit{only sums of odd-order terms occur, while all even order terms in this expression cancel}:

\begin{equation}
\Delta(\Omega)=2\frac{d\phi}{d\Omega}\Omega+\frac{2}{3!}\frac{d^3\phi}{d\Omega^3}\Omega^3+...
\label{Delta}
\end{equation}

Even-order dispersion contributions from the SPDC source will not affect the interference due to the symmetry of signal and idler waves passing through each arm of the interferometer. This phenomenon is similar to other even-order dispersion cancellation effects known in the literature \cite{Franson92,Steinberg92_PRA,Kwiat93,Abouraddy02_PRA}. Note that no dispersion cancellation is present for the phase term $\theta$ introduced by the BDL after NPBS. The expression for $R(y)$ can be written:

\begin{equation}
R(y)=\int_{-\infty}^{\infty}~d\Omega~\left||f(\Omega)|~e^{i\theta}~e^{i\Delta(\Omega)}+|f(-\Omega)|~\right|^2.
\end{equation}

To remove the linear group delay shift of the interferogram pattern, and to concentrate on major interferometric pattern distortions, we compensate the group delay contribution introduced by the source (linear term in Eq. \ref{Delta}) by canceling the group delay in the BDL. In other words we satisfy the following condition $2(d\phi/d\Omega)=\tau_{gr}$. The modified expression for $R(y)$ becomes:

\begin{equation}
R(y)=\int_{-\infty}^{\infty}~d\Omega~\left||f(\Omega)|~e^{i\omega_0\tau_{ph}}~e^{i\delta(\Omega)}+|f(-\Omega)|~\right|^2,
\label{Visibility}
\end{equation}

where $\delta(\Omega)=\Delta(\Omega)-2\frac{d\phi}{d\Omega}\Omega$ is  \textit{the phase term containing only odd dispersion orders $\geq 3$}

\begin{equation}
\delta(\Omega)=\frac{2}{3!}\frac{d^3\phi}{d\Omega^3}\Omega^3+\frac{2}{5!}\frac{d^5\phi}{d\Omega^5}\Omega^5+...~.
\end{equation}

 The behavior of the phase coefficient $\delta(\Omega)$ in conjunction with the joint spectral intensity function $|f(\Omega)|$ is critical for describing complex modulation in the quantum interference function $R(y)$ envelope and for designing a two-photon state that is optimal for high-resolution quantum sensing applications.

\section{Engineering high-quality quantum interference of broadband hyper-entangled states by utilizing  complex  poling profiles in Ti:PPLN}

\subsection{Quantum interference engineering principles}

Some important quantum interference engineering principles emerge from analyzing  Eq.~(\ref{Visibility}).
First, the period of the interference fringe  $\tau_{ph}$  is determined by  $\omega_0\tau_{ph}=2\pi$, or equivalently $\tau_{ph}=2\pi/\omega_0$. It is defined by the optical center frequency of the SPDC spectrum. Second, in order to achieve a $100 \%$ visibility of the envelope one must ensure the coincidence counting rate $R$  vanishes when the average group delay of the source is exactly compensated by the birefringent delay line (BDL). This can only be achieved when a spectral component symmetry condition is fulfilled  $|f(\Omega)|~e^{i\delta}=|f(-\Omega)|$. In other words, two major conditions must be satisfied in the \textit{phase-amplitude parameter space}: i) $|f(\Omega)|=|f(-\Omega)|$ for the amplitude   and ii)  $\delta(\Omega)=0$ for the phase. This is a universal condition of complete indistinguishability between spectral components of correlated signal and idler waves over the full range of nonlinear phase matching in SPDC process.

\subsection{Uniform poling profile}

The entangled photon source we consider here is a quasi-phase matched type-II SPDC process in periodically poled titanium diffused lithium niobate waveguide (Ti:PPLN). Ti:PPLN with a uniform QPM grating period is a standard source of entangled photon pairs for many quantum information and quantum measurement applications. All major spectral and polarization parameters of uniformly poled nonlinear down conversion structure are fully equivalent to those obtained from a bulk crystal. As a result, the "natural" spectral bandwidth of undisturbed uniformly phase-matched source is inversely proportional to the length of nonlinear interaction. Here we consider a $L_{WG}=16.5$~mm long Ti:PPLN structure. The  $\Lambda=9.3~\mu$m  poling period provides phase-matching conditions for the degenerate and collinear SPDC in the case when a 775 nm single longitudinal mode laser is used as a pump and  two 1550 nm correlated daughter photons (signal and idler) are generated.

\begin{figure}[h]
\centering\includegraphics[width=12cm]{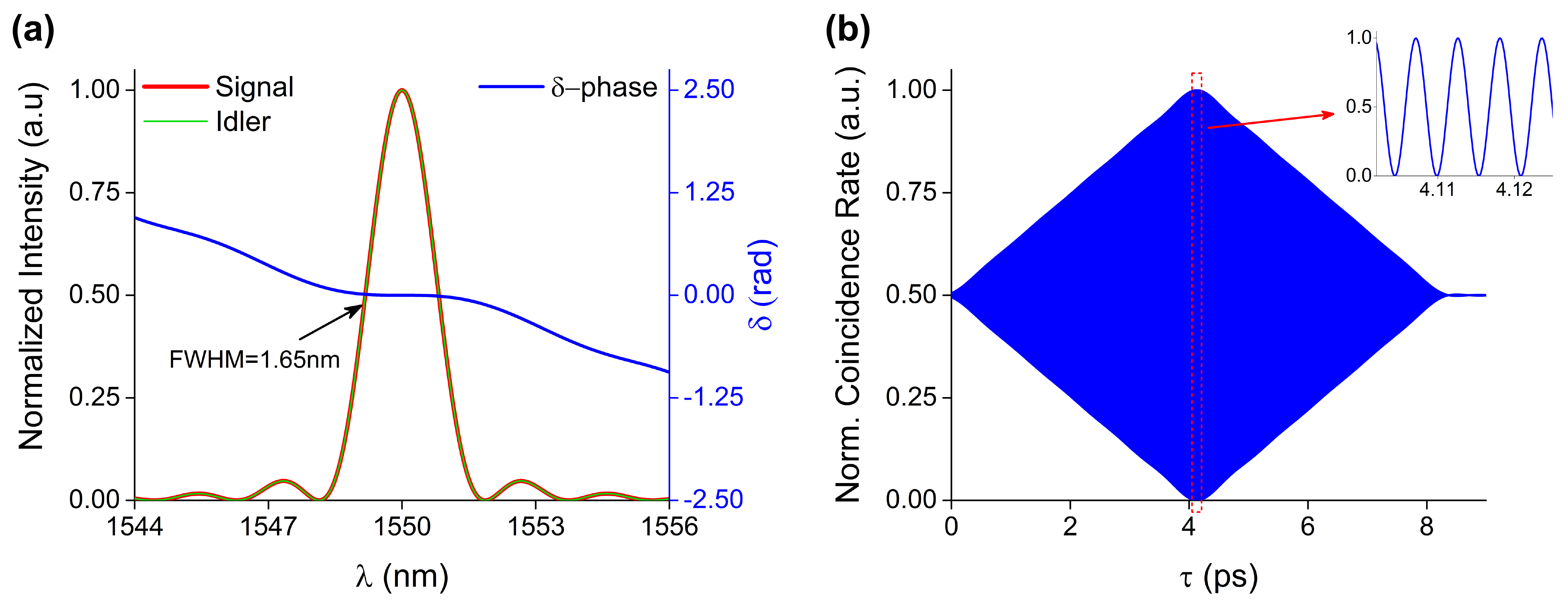}
\caption{Spectral parameters and quantum interference of SPDC radiation obtained in a uniformly poled $16.5$~mm long sample. (a)~perfectly overlapping signal and idler spectral intensity and spectral phase $\delta$ coefficient distribution;  (b)~the intensity correlation interferogram.}
\label{Uniform_Together}
\end{figure}

The spectral intensity of signal and idler radiation, the  distribution of  phase  $\delta(\Omega)$ combining all $\geq 3$-order coefficients, and the quantum interference pattern associated with this signal are shown in Fig.~\ref{Uniform_Together}. The spectral width of the intensity envelope (FWHM) in Fig.~3(a) is 1.65~nm. We scan over the interference fringes by varying the polarization group delay with the BDL. The periodicity of the interference pattern corresponds to a group delay between signal and idler polarized waves with central wavelength of 1550 nm.  The center of the envelope and the maximum of interference visibility occurs at the birefringent delay value $\tau=L_{WG}\left({1}/{u_H}-{1}/{u_V}\right)$ where $u_{H}, (u_{V})$ is the group velocity of the horizontal~(vertical) photon at $\lambda=1550$~nm in the waveguide \cite{Rubin94}.

The triangular shape of this interference pattern is a direct consequence of the $\left(sin(x)/x\right)$  amplitude spectrum of signal and idler waves in the traditional undisturbed parametric down conversion phase-matching. The $\delta$ phase coefficient  is  flat and has zero-value in the vicinity of the intensity peak. The insert shows a high-visibility interference pattern with  a 1550 nm period in the central area, confirming spectral component symmetry. In the complimentary temporal domain this implies that photon pairs generated in the beginning of the nonlinear crystal are indistinguishable from those generated at the end of the nonlinear interaction region.

\subsection{Linear chirping of the poling profile}

\begin{figure}[htbp]
\centering\includegraphics[width=12cm]{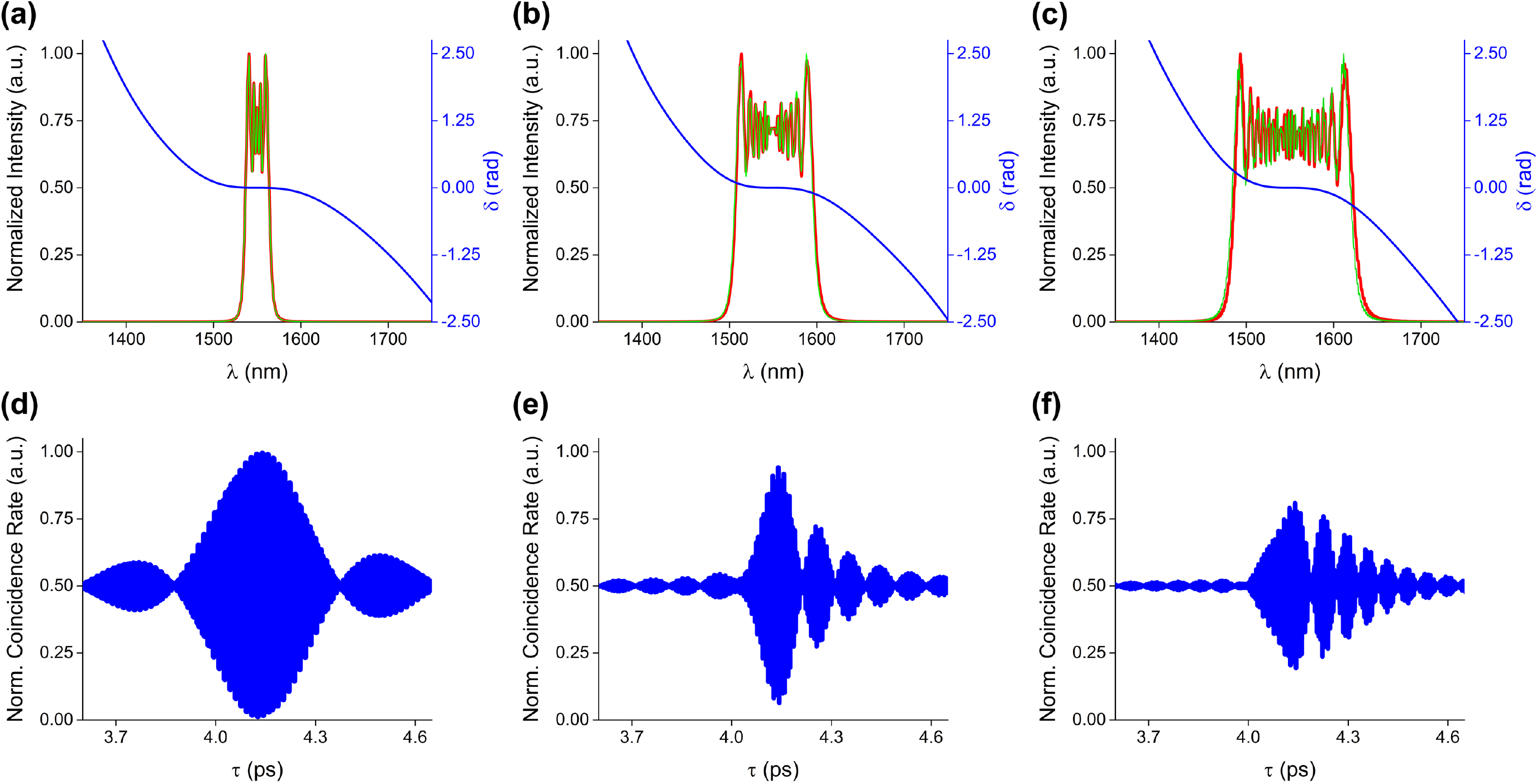}
\caption{The SPDC spectra, $\delta$ phase, and  interferograms  produced by samples with a linear chirp rate of $\alpha=6\times10^{-4}$mm$^{-1}$~(a), $\alpha=18\times10^{-4}$mm$^{-1}$~(b) and $\alpha=30\times10^{-4}$ mm$^{-1}$~(c). The spectral widths of the intensity envelope (FWHM) are $27~$nm~(a), $88$~nm~ (b), and $136$~nm~(c), respectively. The degradation of interference visibility in case of broad spectra obtained with higher chirp rate makes the scheme less useful for interferometric measurement.}
\label{Linear_Chirp}
\end{figure}

The resolution and sensitivity of an interferometric sensor depends on the narrowness of the interference envelope and the peak visibility of the interference fringe. One must broaden the SPDC spectrum in order to reduce the width of the interference envelope and increase the interferometric sensor resolution. It is  known (see for example \cite{Nasr2008}) that one can exploit the linear chirping of the periodic poling profile to broaden the spectral intensity envelope significantly. Consider the poling profile as a function of the length of individual domains $L_m=\Lambda/2~(1+\alpha(z_m-L/2))$, where $\alpha$ is the chirp rate and $L$ is a total length of the nonlinear interaction region. This formula describes a chirping profile with a zero-chirping  point (corresponding to the "natural" uniform period of degenerate SPDC) positioned in the middle of the sample at $L/2$. The resulting SPDC  spectral intensity, the  distribution of   $\delta$ phase, and the associated quantum interference pattern  are shown in Fig.\ref{Linear_Chirp} for three values of chirping rate $\alpha$.

Increasing the chirp rate broadens the SPDC spectrum but also introduces significant distortions in both the intensity spectrum and the interferogram. The width of the main lobe of the interferogram does get smaller following the overall  broadening of the spectrum. However, the significant degradation of the interference visibility and the beating-type distortion of the  envelope limits its usefulness for sensor applications.

The change in the poling profile periodicity (chirp) introduces modulations in the spectral intensity. One way to understand this is that each signal and idler  pair is mapped to a unique nonlinear region around a particular $L_m$ where they are perfectly phase-matched with the highest probability of creation. Only a relatively  small  amount of neighboring domains have poling periods that are very close to what is required for the perfect phase-matching at this wavelength. The varying period results in a shorter effective coherent interaction length for each particular central wavelength, compared to the original total crystal coherent interaction.  The interference envelope modulation and beating in the chirped case occurs because of distinguishability between the interfering amplitudes produced in the beginning and at the end of the nonlinear material, due to significant difference of the poling periods in those areas.  The effect of quantum interference distortion is linked not only to the modulated structure of intensity spectrum (left vertical axis) but also depends on the behavior of higher-order phase  $\delta$ and its deviation from the  $\delta=0$  (right vertical axis) in Fig.~\ref{Linear_Chirp}. The envelope modulation worsens as the chirping parameter $\alpha$ increases and the overall spectrum broadens.

\subsection{Apodization of the linear chirping  profile}

\begin{figure}[htbp]
\centering\includegraphics[width=12	cm]{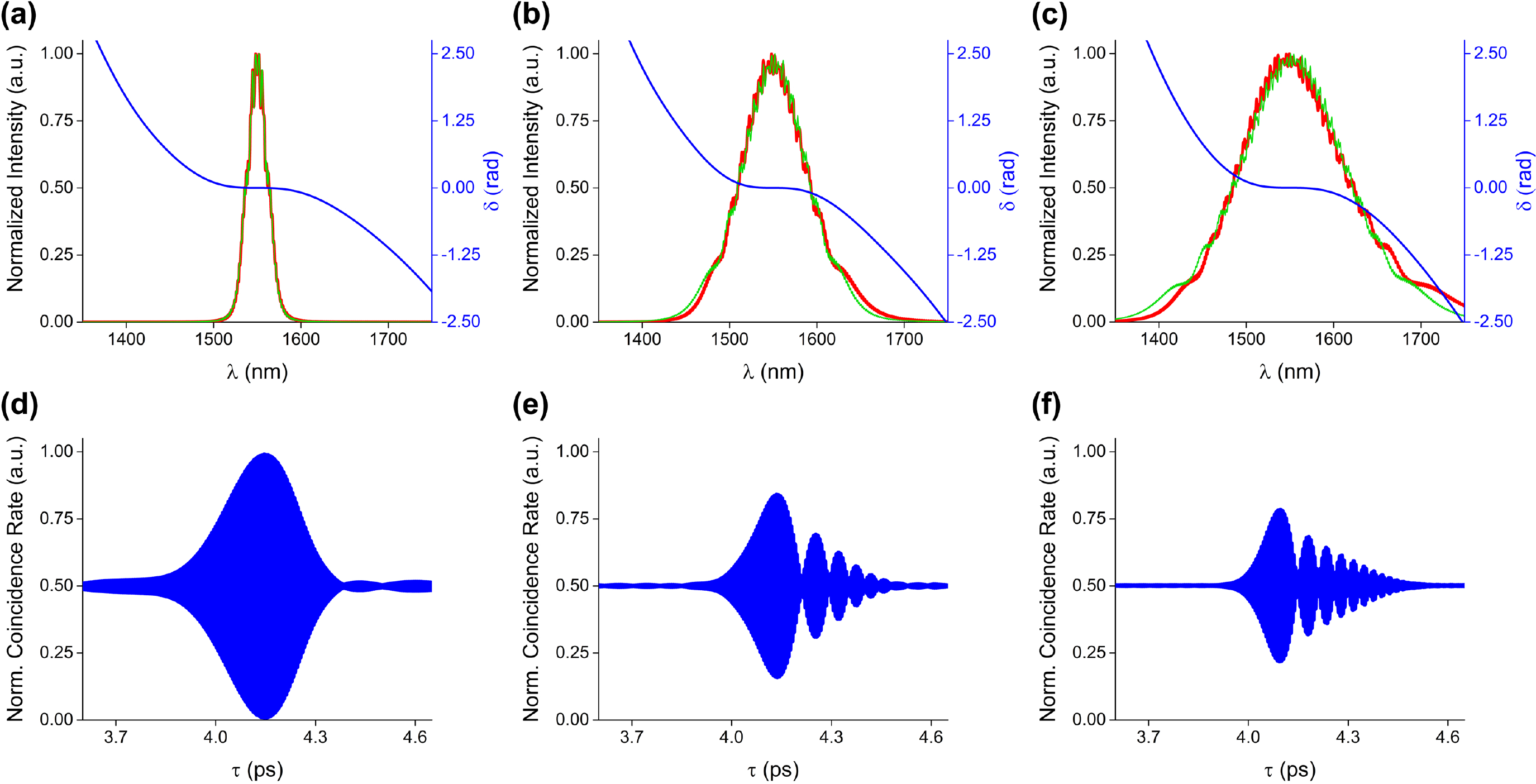}
\caption{The SPDC spectrum, phase $\delta$, and the interferogram produced in a waveguide with apodized poling profile:  (a)~$\alpha=218.9, \beta=-1.99$;  (b)~$\alpha=66.5, \beta=-1.99$; (c)~$\alpha=38.0, \beta=-1.99$. The spectral width of the intensity envelope (FWHM) is $28$~nm~(a), $84$~nm~ (b), and $133$~nm~(c), respectively. The original spectral intensity profile ripples  produced in the linear chirp device (see Fig.\ref{Linear_Chirp}) are smoothed by apodization of the poling profile. The degradation of interference visibility still persists in the case of broad spectra obtained with higher chirp rate.}
\label{Apodized_Linear_Chirp}
\end{figure}

The problem of symmetrizing the spectral intensity distribution and satisfying the condition $|f(\Omega)|=|f(-\Omega)|$  was addressed in the literature  during the design of efficient optical parametric power converters  \cite{HeeseOptExpress2012}.   A traditional way to smooth the broadband phase-matching  and restore a Gaussian spectral intensity distribution is to apodize the linear chirp of the periodic poling profile. The idea is to change the poling period more rapidly at the edges of the device while keeping the monotonic linear chirp in the middle. Specifically, we consider a poling profile of the form $L_m=\Lambda/2~(1+\frac{1}{\alpha}~atanh(\beta(z_m-L/2)/L))$. The resulting spectra are shown in Fig.~\ref{Apodized_Linear_Chirp}. The process of apodization clearly smooths the intensity spectrum envelope and makes it more symmetric while preserving the enlarged FWHM bandwidth.

However, examination of the interference pattern in Fig.~\ref{Apodized_Linear_Chirp} reveals that the interferogram quality still suffers significantly from  chirping.  The problem lies in the second indistinguishability criterion that is linked with the behavior of higher-order phase coefficient  $\delta$.

\begin{figure}[htbp]
\centering\includegraphics[width=8 cm]{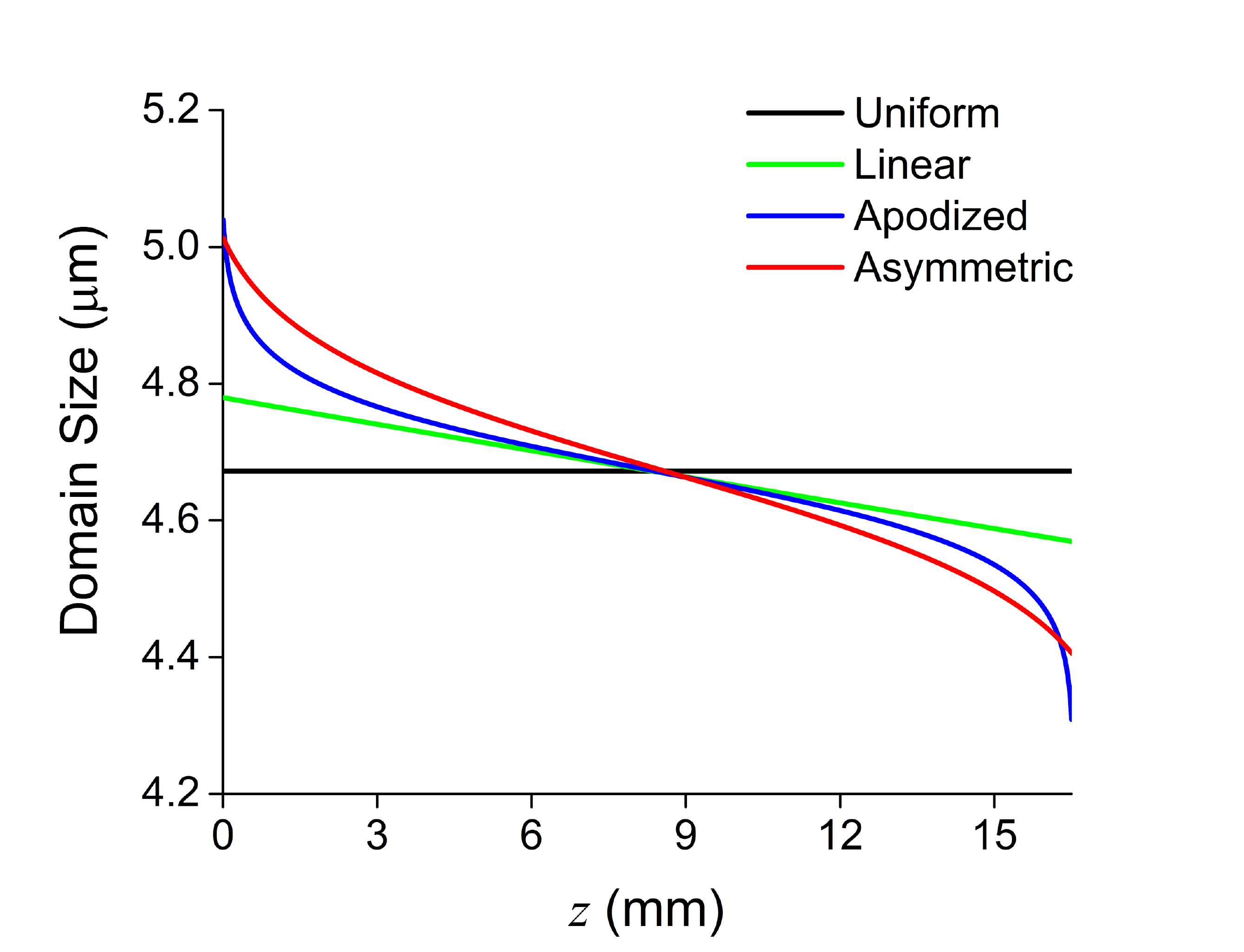}
\caption{Different poling  profiles along the length of nonlinear crystal interaction region $z$. }
\label{Evolution}
\end{figure}

\section{New  approach - asymmetric poling of chirped Ti:PPLN}

We  established that a flat zero-value spectrum of residual phase $\delta$  and symmetric intensity spectra for signal and idler frequencies are required conditions to ensure good visibility in quantum interferometry. Furthermore,  a broadband spectrum is needed to achieve a narrow interferogram. All previous poling approaches utilized uniform nonlinear poling, chirped poling, or apodization poling profiles that were symmetric relative to the middle point of the crystal in order to preserve the timing indistinguishability of the generated entangled photons (see Fig.~\ref{Evolution}). We introduce  here a new approach that abandons this  artificially imposed  "spatial central point" symmetry condition in favor of a more general alternative that symmetrizes components of the generated state in the phase-amplitude parameter space.

 We designed the following polynomial poling profile for this purpose:
\begin{equation}
 L_m=\Lambda/2~(1+\sum_{j=0}^{N}\alpha_{2j+1}(z_m-\beta_{2j+1}L/2)^{2j+1}.
\end{equation}

Unlike the previously described chirped or apodized profiles, the general profile is now made asymmetric (see Fig.~\ref{Evolution}). This is achieved by allowing the center of the j-th order in the profile to be offset from the center of the interaction region by a factor $\beta_j$. We selected the odd-order polynomial structure because it fits better with the strategy of asymmetric poling optimization. Note that this profile coincides with a linearly chirped case when $N=0$. The strength of each order $\alpha_j$ and the asymmetry factor $\beta_j$ are both optimized by utilizing a genetic algorithm~(GA) convergence procedure for different values of $N$.

Based on this approach we introduced a  target function~(G) for the optimization process:

\begin{equation}
G_1(\sigma)=\int{d\Omega}\left[|\delta(\Omega)|^2+ ||f(\Omega)|-|f(-\Omega)||^2\right]+|FWHM\{|f(\Omega)|^2\}-\sigma|^2.
\end{equation}

\begin{figure}[htbp]
\centering\includegraphics[width=12 cm] {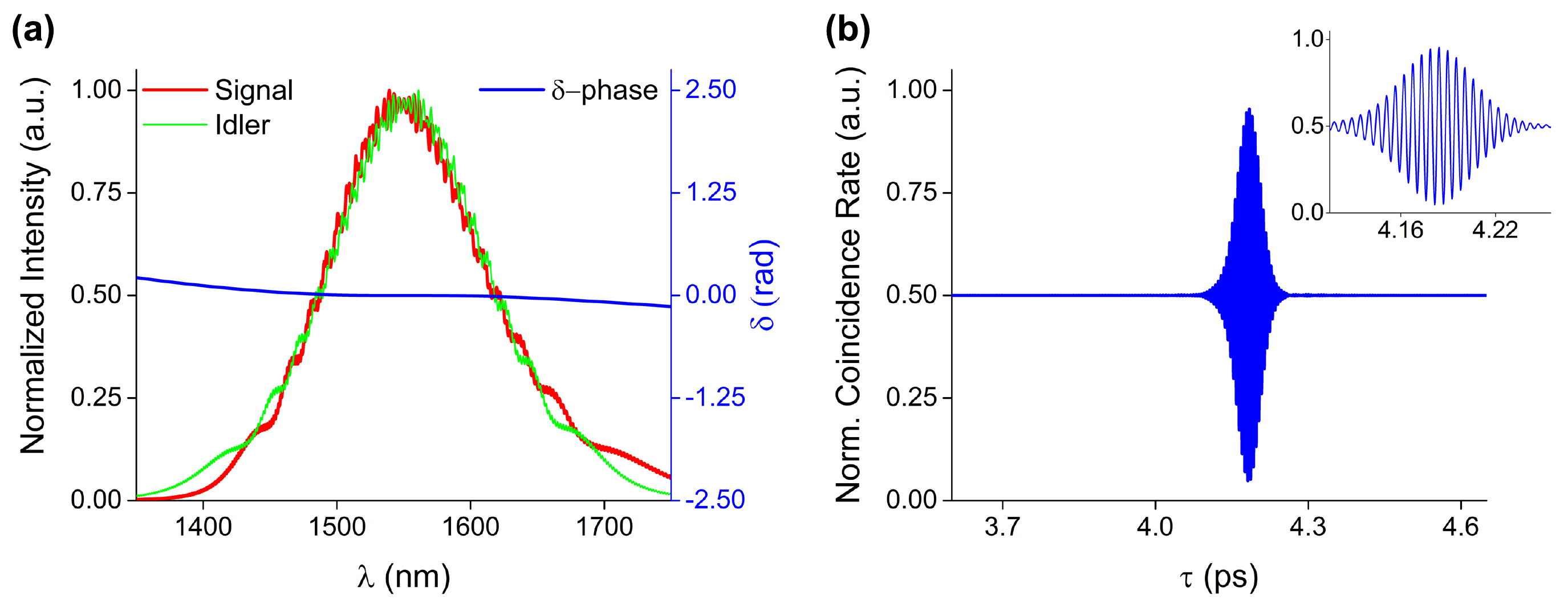}
\caption{The SPDC spectrum, phase $\delta$, and the optimized interferogram produced using asymmetric poling Eq.(15) in case of $N=11$. The spectral width of the intensity envelope (FWHM) is $135$~nm.}
\label{Poling_Asymmetric}
\end{figure}

The first term is a least-squares minimization of the residual phase $\delta(\Omega)$. The second term minimizes the asymmetry between correlated signal and idler spectral amplitudes in each photon pair. The third term is designed to achieve a specified full-width at half maximum~(FWHM) bandwidth of the spectral amplitude.  The optimization procedure implements a constraint to keep the minimum domain size to be above our lithographically feasible minimum domain size of $4~\mu$m. The GA is chosen as our optimization technique because the problem of reaching the target function with an arbitrary poling profile is not well suited for optimization algorithms based on numerical differentiation. The initial step of the algorithm chooses arbitrary values of  parameters $\alpha_j$ and $\beta_j$ conditioned with the minimum domain size constraint. The resulting interferograms obtained using optimization procedure Eq.~(14) in case of  $N=11$ is shown in Fig.~\ref{Poling_Asymmetric}. The optimized asymmetric poling profile is illustrated in Fig.~\ref{Evolution}.

\section{Discussion}

The main result  (see Fig.~\ref{Poling_Asymmetric})  illustrates that the method proposed here, of asymmetric  chirped poling structure in the Ti-diffused nonlinear waveguide, is capable of simultaneously satisfying two conditions: i) significantly increasing the bandwidth of entangled states and ii) ensuring amplitude and phase indistinguishability of signal and idler spectral components. The resulting interferogram is a clean single peak with a  narrow temporal envelope of 54 fs FWHM and 90.3\% visibility.

The approach presented here of two-photon quantum optical interferometry using a specially engineered broadband source of hyper-entangled states becomes an exceptional tool in interferometric sensing of group delays, because it enjoys several special features not present in classical interferometry.

One interesting feature of this quantum interference arrangement is the presence of the zero-order dispersion cancellation. The cancellation of the $\phi(0)$ term in Eq.~(\ref{Visibility}) (along with the second-order dispersion and following even orders) has an important physical consequence: a phase velocity interference fringe  $\tau_{ph}=2\pi/\omega_0$ of the degenerate down conversion $\omega_0=\omega_p/2$  is not affected by the sample properties (see \cite{fraine2011precise}). In other words, the fringe location is not sensitive to the presence of the sample. This offers a natural phase delay reference frame because the phase of the interferogram remains undisturbed and interference fringes do not shift. This is in strong contrast with other types of conventional phase interference. The value of the group delay (first-order dispersion term) experienced by the light in the sample is now evaluated by observing how the relatively narrow group envelope is sliding along the stationary phase modulation pattern. The effect can be viewed conceptually as a presence of fixed phase delay ruler that is not altered by the object under investigation.  The main benefit is  that the phase interference is capable of detecting much smaller delay values when compared to the group delay evaluation.  The simultaneous presence  of both patterns in the interferogram - stationary phase delay ruler and movable group delay envelope - can enhance the resolution of group delay evaluation to the same accuracy level as phase delay measurements.

In addition,  differential loss does not degrade the interference pattern quality. If one photon of the pair is lost (absorbed or scattered out of the optical path), no coincidence occurs. The second photon may be detected, but the pair will not be recorded. Therefore, only the rate of  coincidence counting  will be affected by the differential loss, but it will not reduce the visibility or resolution of the interferometric sensor. 

The reported asymmetric poling profile in Fig.~\ref{Evolution} leads to the  generation of  broadband hyper-entangled entangled states as illustrated in Fig.~\ref{Poling_Asymmetric}. One practical advantage of fabricating asymmetric poling profiles is based on a more mild domain size change at the crystal edges in comparison with a standard apodization procedure (see Fig.~\ref{Evolution}). Experimental realization of such a device would enable all of the above mentioned  quantum interference features and create a super-resolution sensor when the group delay will be evaluated with the resolution reserved traditionally only for the phase measurement.

\section{Conclusion}

We designed a broadband source of polarization and frequency entangled photon pairs for quantum interferometry using nonlinear parametric down conversion in a strongly chirped Ti-diffused PPLN. The approach for designing  indistinguishable signal and idler  waves  in a wide range of frequencies is based on asymmetric poling that equalizes signal and idler components in the magnitude and phase space instead of conventional practice of ensuring a symmetric place of origin in the nonlinear crystal. The produced state will be able to offer a  narrow  interferogram without any beating features.  This high quality of interference pattern and  nonclassical dispersion cancellation effects will be crucial for building high-resolution sensors.

\end{document}